\documentclass[10pt]{article}

\usepackage{graphicx}%
\usepackage{multirow}%
\usepackage{amsmath,amssymb,amsfonts}%
\usepackage{amsthm}%
\usepackage{mathrsfs}%
\usepackage{xcolor}%
\usepackage{textcomp}%
\usepackage{manyfoot}%
\usepackage{booktabs}%
\usepackage{algorithm}%
\usepackage{algorithmicx}%
\usepackage{algpseudocode}%
\usepackage{listings}%

\usepackage{bbm}

\usepackage[margin=1.4in]{geometry} % full-width

\newcommand{\bfx}{{\bf x}}
\newcommand{\bfr}{{\bf r}}

\title{A note on bounded distance-based information loss metrics for statistical disclosure control of numeric microdata}

\author{Elias Chaibub Neto}

\date{Sage Bionetworks, Seattle, WA 98121, USA \\ elias.chaibub.neto@sagebase.org}

\begin{document}

\maketitle

\begin{abstract}
In the field of statistical disclosure control, the tradeoff between data confidentiality and data utility is measured by comparing disclosure risk and information loss metrics. Distance based metrics such as the mean absolute error (MAE), mean squared error (MSE), mean variation (IL1), and its scaled alternative (IL1s) are popular information loss measures for numerical microdata. However, the fact that these measures are unbounded makes it is difficult to compare them against disclosure risk measures which are usually bounded between 0 and 1. In this note, we propose rank-based versions of the MAE and MSE metrics that are bounded in the same range as the disclosure risk metrics. We empirically compare the proposed bounded metrics against the distance-based metrics in a series of experiments where the metrics are evaluated over multiple masked datasets, generated by the application of increasing amounts of perturbation (e.g., by adding increasing amounts of noise). Our results show that the proposed bounded metrics produce similar rankings as the traditional ones (as measured by Spearman correlation), suggesting that they are a viable additions to the toolbox of distance-based information loss metrics currently in use in the SDC literature. \\

\end{abstract}

\section{Introduction}

The goal of statistical disclosure control (SDC) is to generate masked datasets that minimize the risk of giving away confidential information while, at the same time, retaining as much analytical utility as possible~\cite{Drechsler2011,hundepool2012}. Masked datasets are usually generated via an iterative process where small perturbations are applied to data, which is then evaluated to make sure that confidentiality and utility are maintained at acceptable levels. During this process, the tradeoff between data confidentiality and data utility is traditionally measured using a combination of disclosure risk (DR) and information loss (IL) metrics~\cite{DomingoFerrerTorra2001}. The use of multiple metrics is generally preferred because different metrics capture different aspects of the similarity between the original and masked datasets.

Quite importantly, because widely used DR metrics such as distance-based record linkage~\cite{PagliucaSeri1999,DomingoFerrerTorra2001}, rank-based interval distance~\cite{DomingoFerrerTorra2001}, and standard deviation-based interval distance~\cite{MateoSanzDomingoSebeFerrer2004}, are bounded between 0 and 1, there has been a lot of effort in the development of IL measures bounded in the same range, in order to facilitate the comparison with the DR metrics. Examples include the probabilistic information loss metric~\cite{PilPaper2005}, the propensity score based IL metric~\cite{Woo2009} (which is actually bounded between [0, 0.25]), and the covariance based bounded IL metric~\cite{DomingoFerrerMuralidharBrasAmoros2020}. However, popular distance based information loss metrics such as the mean absolute error (MAE)~\cite{DomingoFerrerMateoSanzTorra2001,DomingoFerrerTorra2001b}, the mean squared error (MSE)~\cite{DomingoFerrerMateoSanzTorra2001,DomingoFerrerTorra2001b}, the mean variation (IL1)~\cite{MateoSanzDomingoSebeFerrer2004}, and its more stable scaled alternative (IL1s)~\cite{Yancey2002,MateoSanzDomingoSebeFerrer2004}, are unbounded.

In this note, we propose rank-based versions of the MAE and MSE metrics, which are bounded between 0 and 1. We denoted these new metrics by, brMAE and brMSE. Intuitively, because the ranks of a variable are always bounded between 1 and the number of records, $n$,\footnote{Strictly speaking, the ranks will be bounded between 1 and $n$ if there are no ties in the data. However, if ties are present, they can always be broken using a method that does not generate repeated values such as random assignment of ranks to the tied values, or the assignment of increasing (or decreasing) values at each index set of ties.} it is easy to compute the maximal possible distance between a variable in the original and masked datasets. Then, by dividing the unbounded rank-based distance metric by its maximal distance we obtain a bounded metric in [0, 1] interval.

While working with ranks rather than with the raw numerical values might potentially lead to some loss of information (in the sense that we reduce the granularity with which we can measure the distance between the original and masked datasets) the ability to obtain a bounded distance metric might, in itself, out-weight this limitation.

In any case, in order to assess the potential impact of this limitation, we performed empirical comparisons between the proposed brMAE and brMSE metrics against the MAE, MSE, IL1 and IL1s metrics. In our comparisons we used popular perturbation methods, namely, microaggregation~\cite{DomingoFerrerMateoSanz2002}, noise addition~\cite{Brand2004}, and rank swapping~\cite{Moore1996}, applied to 2 datasets traditionally used for the comparison of numeric microdata (namely, the Tarragona and Census datasets, which are publicly available with the sdcMicro R package~\cite{sdcMicro2015}). Each perturbation method was used to produce a series of masked datasets generated by the application of increasing amounts of perturbation. Our empirical results show that the rankings computed with the bounded distance metrics (brMAE and brMSE) tend to correlate very well with the rankings obtained with the unbounded ones (MAE, MSE, IL1, and IL1s).

\section{The proposed new metrics}

Let $\bfx = (x_{1}, x_{2}, \ldots, x_{n})^T$ and $\tilde{\bfx} = (\tilde{x}_{1}, \tilde{x}_{2}, \ldots, \tilde{x}_{n})^T$ represent, respectively, the original and perturbed raw data vectors while $\bfr = (r_{1}, r_{2}, \ldots, r_{n})^T$ and $\tilde{\bfr} = (\tilde{r}_{1}, \tilde{r}_{2}, \ldots, \tilde{r}_{n})^T$ represent their respective ranks. Assume there are no ties so that for both vectors the ranks range from 1 to $n$ by increments of 1. (If there are ties in the data, the ties should be broken by methods that generate unique rank values, such as, for example, random assignment of ranks to the tied values.)

Because both $\bfr$ and $\tilde{\bfr}$ vectors have length $n$ and have no repeats, it is clear that the maximum possible absolute difference between the values at any position of these vectors is $n - 1$ (i.e., when there exists a position $i$ for which either $r_{i} = n$ and $\tilde{r}_i = 1$ or $r_{i} = 1$ and $\tilde{r}_i = n$). If there exists one position $i$ for which the maximum absolute difference is $n - 1$, then the second maximum possible absolute difference is given by $n - 3$ (i. e., when there exists another position $i'$ for which either $r_{i'} = n - 1$ and $\tilde{r}_{i'} = 2$ or $r_{i'} = 2$ and $\tilde{r}_{i'} = n - 1$, since the ranks $n$ and 1 where already taken by position $i$). If there exists positions $i$ and $i'$ for which the maximum and second maximum absolute differences are $n - 1$ and $n - 3$, then the next maximum possible difference is given by $n - 5$, and so on.

To formalize the above argument consider, without loss of generality, two vectors, $\bfr_1^\star$ and $\bfr_2^\star$, such that $\bfr_1^\star = (1, 2, \ldots, n-1, n)^T$ and $\bfr_2^\star = (n, n-1, \ldots, 2, 1)^T$ (i.e., $\bfr_2^\star$ is the reverse of $\bfr_1^\star$). Clearly, no other two rank vectors can be farther away from each other in terms of the sum of the absolute (or squared) distances between their elements. Note that if $n$ is even,
$$
\bfr_1^\star = \left(\hspace{-0.2cm}
\begin{array}{c}
1 \\
2 \\
3 \\
\vdots \\
\frac{n}{2} \\
\frac{n}{2} + 1 \\
\vdots \\
n - 2 \\
n - 1 \\
n \\
\end{array}
\hspace{-0.2cm} \right)~,
\hspace{0.2cm}
\bfr_2^\star = \left(\hspace{-0.2cm}
\begin{array}{c}
n \\
n - 1 \\
n - 2 \\
\vdots \\
\frac{n}{2} + 1 \\
\frac{n}{2} \\
\vdots \\
3 \\
2 \\
1 \\
\end{array}
\hspace{-0.2cm} \right)~,
\hspace{0.2cm}
\bfr_1^\star - \bfr_2^\star = \left(\hspace{-0.2cm}
\begin{array}{c}
1 - n \\
2 - (n - 1) \\
3 - (n - 2) \\
\vdots \\
\frac{n}{2} - (\frac{n}{2} + 1)  \\
\frac{n}{2} + 1 - \frac{n}{2} \\
\vdots \\
(n - 2) - 3 \\
(n - 1) - 2 \\
n - 1 \\
\end{array}
\hspace{-0.2cm} \right)
=
\left(\hspace{-0.1cm}
\begin{array}{c}
1 - n \\
3 - n \\
5 - n \\
\vdots \\
-1 \\
1 \\
\vdots \\
n - 5 \\
n - 3 \\
n - 1 \\
\end{array}
\hspace{-0.1cm} \right)~,
$$
and we have that the total absolute distance between $\bfr_1^\star$ and $\bfr_2^\star$ is given by,
\begin{align}
\sum_{i = 1}^n |r_{1i}^\star - r_{2i}^\star| &= |1 - n| + |3 - n| + \ldots + |-1| + |1| + \ldots + |n - 3| + |n - 1| \nonumber \\
&= 2 \, (n - 1) + 2 \, (n - 3) + \ldots + 2 \, (1) \, = \, 2 \, \sum_{k = 1}^{n/2} (n - 2 \, k + 1)~.
\end{align}
Similarly, in the case that $n$ is odd, we have,
$$
\bfr_1^\star = \left(\hspace{-0.2cm}
\begin{array}{c}
1 \\
2 \\
3 \\
\vdots \\
\frac{n - 1}{2} \\
\frac{n - 1}{2} + 1 \\
\frac{n - 1}{2} + 2 \\
\vdots \\
n - 2 \\
n - 1 \\
n \\
\end{array}
\hspace{-0.2cm} \right),
\hspace{0.1cm}
\bfr_2^\star = \left(\hspace{-0.2cm}
\begin{array}{c}
n \\
n - 1 \\
n - 2 \\
\vdots \\
\frac{n - 1}{2} + 2 \\
\frac{n - 1}{2} + 1 \\
\frac{n - 1}{2} \\
\vdots \\
3 \\
2 \\
1 \\
\end{array}
\hspace{-0.2cm} \right),
\hspace{0.1cm}
\bfr_1^\star - \bfr_2^\star = \left(\hspace{-0.2cm}
\begin{array}{c}
1 - n \\
2 - (n - 1) \\
3 - (n - 2) \\
\vdots \\
\frac{n - 1}{2} - (\frac{n - 1}{2} + 2)  \\
\frac{n - 1}{2} + 1 - (\frac{n - 1}{2} + 1) \\
\frac{n - 1}{2} + 2 - (\frac{n - 1}{2}) \\
\vdots \\
(n - 2) - 3 \\
(n - 1) - 2 \\
n - 1 \\
\end{array}
\hspace{-0.2cm} \right)
=
\left(\hspace{-0.1cm}
\begin{array}{c}
1 - n \\
3 - n \\
5 - n \\
\vdots \\
-2 \\
0 \\
2 \\
\vdots \\
n - 5 \\
n - 3 \\
n - 1 \\
\end{array}
\hspace{-0.1cm} \right),
$$
and the total absolute distance is given by,
\begin{align}
\sum_{i = 1}^n |r_{1i}^\star - r_{2i}^\star| &= |1 - n| + |3 - n| + \ldots + |-2| + |0| + |2| + \ldots + |n - 3| + |n - 1| \nonumber \\
&= 2 \, (n - 1) + 2 \, (n - 3) + \ldots + 2 \, (2) + 0 \, = \, 2 \, \sum_{k = 1}^{(n - 1)/2} (n - 2 \, k + 1)~.
\end{align}

Similarly, the total squared distance between $\bfr_1^\star$ and $\bfr_2^\star$ is given by,
\begin{equation}
\sum_{i = 1}^n (r_{1i}^\star - r_{2i}^\star)^2 \, = \, 2 \, \sum_{k = 1}^{K} (n - 2 \, k + 1)^2~.
\end{equation}
where $K$ is equal to $n/2$ when $n$ is even, and to $(n-1)/2$ when it is odd.

Here, it is important to clarify that while the absolute distance between $\bfr_1^\star$ and $\bfr_2^\star$ is maximal, there are other possible configurations of rank vectors that also achieve the same maximal distance (but cannot surpass it). As an illustration, consider all the possible permutations of rank vectors of dimension $n = 3$ and $n = 4$ presented on Tables \ref{tab:1} and \ref{tab:2}, respectively. When $n = 3$ (Table \ref{tab:1}) there are $n! = 6$ possible permutations of the $\bfr_1^\star = (1, 2, 3)^T$ vector, denoted by $\bfr_2$, and reported in rows 2 to 4 of Table \ref{tab:1}. (Note that permutation number 1 corresponds to $\bfr_1^\star$ while permutation number 6 corresponds to $\bfr_2^\star = (3, 2, 1)^T$.) Rows 5 to 7 show $\bfr_1^\star - \bfr_2$ for all six possible permutations, while rows 8 and 9 present, respectively, the absolute and the squared distances between the $\bfr_1^\star$ and $\bfr_2$ vectors. Observe that, in addition to permutation number 6, permutations number 4 and 5 also achieve the maximal possible absolute distance value of 4.

\begin{table}[!h]
\caption{Example for $n = 3$}
  \label{tab:1}
  \centering
{\footnotesize
\setlength\tabcolsep{2.5pt}
\begin{tabular}{l|rrrrrrrrrrrrrrrrrrrrrrrr}
\toprule
permutation & 1 & 2 & 3 & 4 & 5 & 6 \\
\midrule
& 1 & 1 & 2 & 3 & 2 & 3 \\
$\bfr_2$ & 2 & 3 & 1 & 1 & 3 & 2 \\
& 3 & 2 & 3 & 2 & 1 & 1 \\
\midrule
& 0 & 0 & -1 & -2 & -1 & -2 \\
$\bfr_1^\star - \bfr_2$ & 0 & -1 & 1 & 1 & -1 & 0 \\
& 0 & 1 & 0 & 1 & 2 & 2 \\
\midrule
$\sum_{i = 1}^3 |r_{1i}^\star - r_{2i}|$ & 0 & 2 & 2 & 4 & 4 & 4 \\
$\sum_{i = 1}^3 (r_{1i}^\star - r_{2i})^2$ & 0 & 2 & 2 & 6 & 6 & 8 \\
\bottomrule
\end{tabular}}
\end{table}

When $n = 4$ (Table \ref{tab:2}) there are $n! = 24$ possible permutations of the $\bfr_1^\star = (1, 2, 3, 4)^T$ vector. In this case, in addition to permutation number 24, permutations number 20, 21, and 22 also achieve the maximal possible absolute distance value of 8. Observe that in addition to having multiple permutations achieving the maximal distance, the absolute rank distance tends to produce a less granular range of values than the squared rank distance. (Note that the absolute distance only assumes five distinct values, namely, 0, 2, 4, 6, and 8, while the squared distance assumes 10 distinct values, namely, 0, 2, 4, 6, 8, 10, 12, 14, 16, 18, and 20.)

\begin{table}[!h]
\caption{Example for $n = 4$}
  \label{tab:2}
  \centering
{\footnotesize
\setlength\tabcolsep{2.5pt}
\begin{tabular}{l|rrrrrrrrrrrrrrrrrrrrrrrr}
\toprule
permutation & 1 & 2 & 3 & 4 & 5 & 6 & 7 & 8 & 9 & 10 & 11 & 12 & 13 & 14 & 15 & 16 & 17 & 18 & 19 & 20 & 21 & 22 & 23 & 24 \\
\midrule
& 1 & 1 & 1 & 2 & 2 & 1 & 1 & 3 & 2 & 1 & 3 & 3 & 2 & 4 & 2 & 4 & 3 & 2 & 4 & 3 & 4 & 3 & 4 & 4 \\
$\bfr_2$ & 2 & 2 & 3 & 1 & 1 & 4 & 3 & 1 & 3 & 4 & 2 & 1 & 4 & 1 & 3 & 1 & 2 & 4 & 2 & 4 & 3 & 4 & 2 & 3 \\
& 3 & 4 & 2 & 3 & 4 & 2 & 4 & 2 & 1 & 3 & 1 & 4 & 1 & 2 & 4 & 3 & 4 & 3 & 1 & 1 & 1 & 2 & 3 & 2 \\
& 4 & 3 & 4 & 4 & 3 & 3 & 2 & 4 & 4 & 2 & 4 & 2 & 3 & 3 & 1 & 2 & 1 & 1 & 3 & 2 & 2 & 1 & 1 & 1 \\
\midrule
& 0 & 0 & 0 & -1 & -1 & 0 & 0 & -2 & -1 & 0 & -2 & -2 & -1 & -3 & -1 & -3 & -2 & -1 & -3 & -2 & -3 & -2 & -3 & -3 \\
$\bfr_1^\star - \bfr_2$ & 0 & 0 & -1 & 1 & 1 & -2 & -1 & 1 & -1 & -2 & 0 & 1 & -2 & 1 & -1 & 1 & 0 & -2 & 0 & -2 & -1 & -2 & 0 & -1 \\
& 0 & -1 & 1 & 0 & -1 & 1 & -1 & 1 & 2 & 0 & 2 & -1 & 2 & 1 & -1 & 0 & -1 & 0 & 2 & 2 & 2 & 1 & 0 & 1 \\
& 0 & 1 & 0 & 0 & 1 & 1 & 2 & 0 & 0 & 2 & 0 & 2 & 1 & 1 & 3 & 2 & 3 & 3 & 1 & 2 & 2 & 3 & 3 & 3 \\
\midrule
$\sum_{i = 1}^4 |r_{1i}^\star - r_{2i}|$ & 0 & 2 & 2 & 2 & 4 & 4 & 4 & 4 & 4 & 4 & 4 & 6 & 6 & 6 & 6 & 6 & 6 & 6 & 6 & 8 & 8 & 8 & 6 & 8 \\
$\sum_{i = 1}^4 (r_{1i}^\star - r_{2i})^2$ & 0 & 2 & 2 & 2 & 4 & 6 & 6 & 6 & 6 & 8 & 8 & 10 & 10 & 12 & 12 & 14 & 14 & 14 & 14 & 16 & 18 & 18 & 18 & 20 \\
\bottomrule
\end{tabular}}
\end{table}

\begin{figure}[!h]
\centerline{\includegraphics[width=\linewidth]{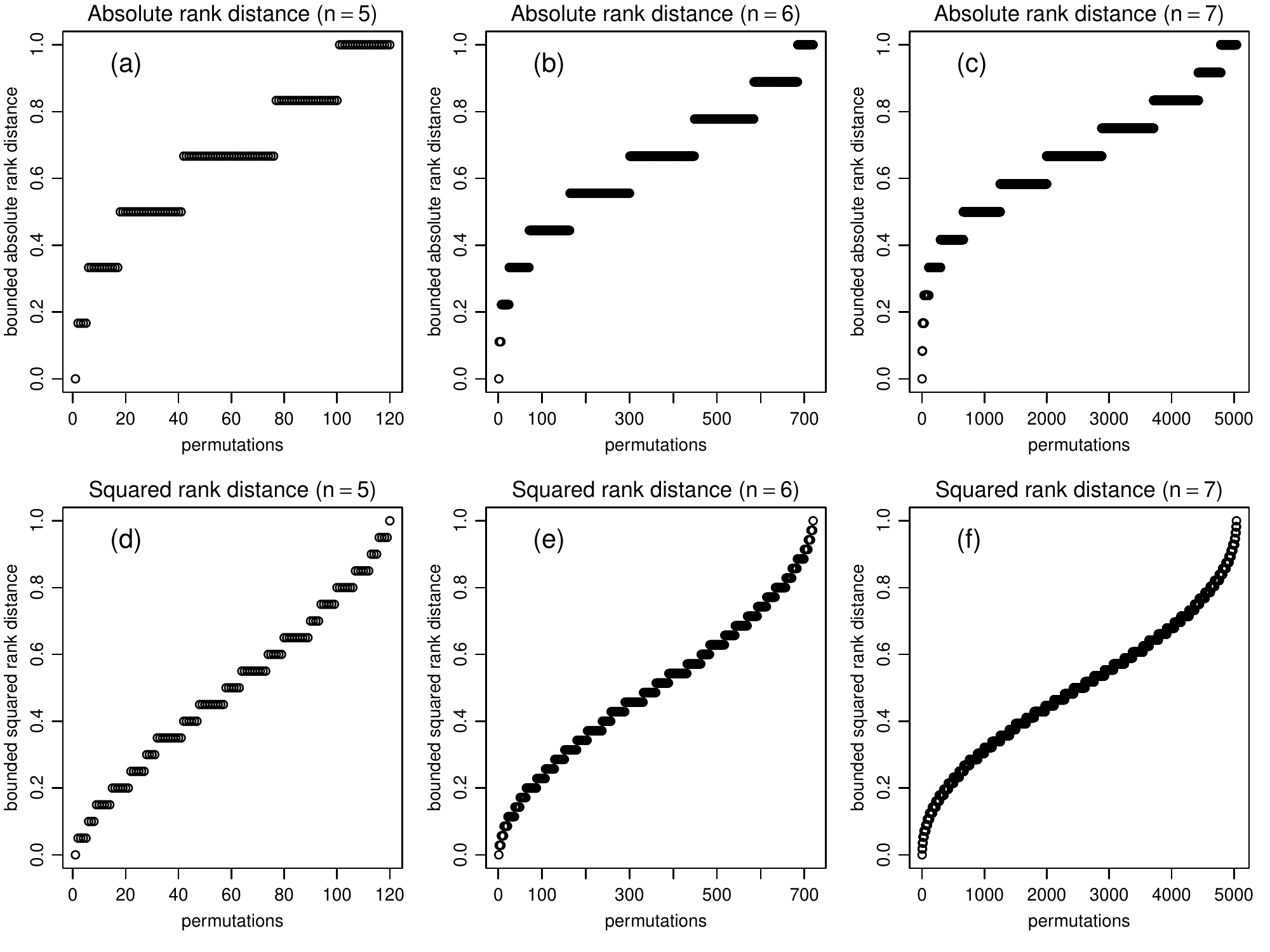}}
%\vskip -0.1in
\caption{Bounded absolute and squared distances for $n = 5$, $n = 6$, and $n = 7$.}
\label{fig:bounded.distances.5.6.7}
%\vskip -0.2in
\end{figure}

Figure \ref{fig:bounded.distances.5.6.7} presents bounded values of the absolute and squared distances between $\bfr_1^\star$ and $\bfr_2$ for $n = 5$, $n = 6$, and $n = 7$, where the bounded absolute and square rank distances are computed, respectively, as,
\begin{equation}
\sum_{i = 1}^{n} |r_{1i}^\star - r_{2i}|/\sum_{k = 1}^{K} 2 \, (n - 2 \, k + 1)~,
\end{equation}
and
\begin{equation}
\sum_{i = 1}^{n} (r_{1i}^\star - r_{2i})^2/\sum_{k = 1}^{K} 2 \, (n - 2 \, k + 1)^2~.
\end{equation}
Each panel in Figure \ref{fig:bounded.distances.5.6.7} reports the bounded rank distance (y-axis) for all possible permutations of the rank vectors of length $n$ (x-axis). (For $n$ equal to 5, 6, and 7, there are $n!$ equal to 120, 720, and 5,040 possible permutations.) Panels a, b, and c report the absolute rank distances, while panels d, e, and f report the squared rank distances. The top panels in Figure \ref{fig:bounded.distances.5.6.7} show, once again, that multiple rank permutations are able to achieve the maximal absolute distance of $\sum_{k = 1}^{K} 2 \, (n - 2 \, k + 1)$. Observe, as well, that the absolute rank distance again tends to produce a less granular range of values than the squared rank distance (note the smaller number of steps across the y-axis of the top panels in comparison with the bottom panels).

\begin{figure}[!h]
\centerline{\includegraphics[width=4in]{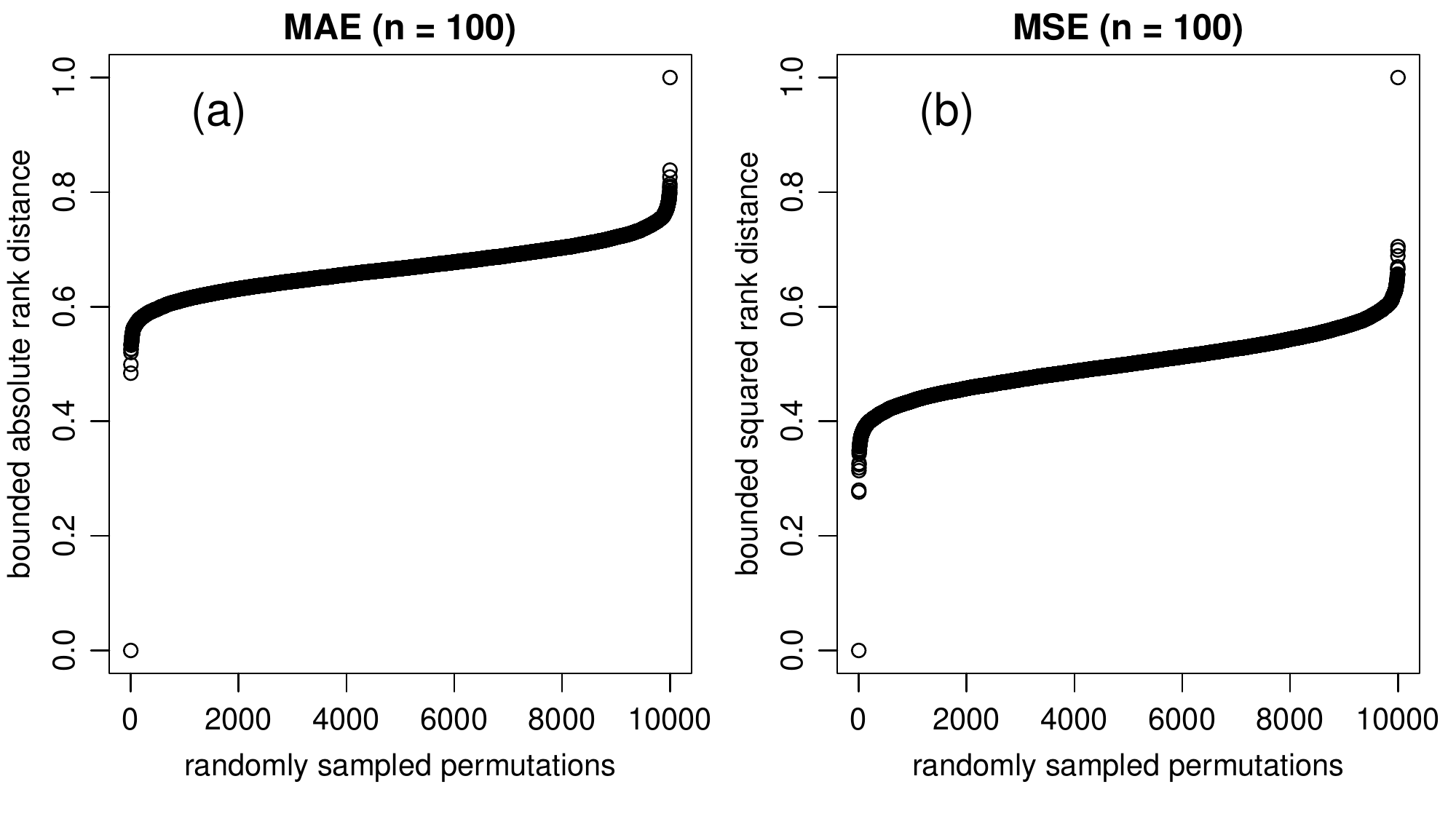}}
%\vskip -0.1in
\caption{Bounded absolute and squared distances for $n = 100$, from 10,000 random permutations.}
\label{fig:bounded.distances.100}
%\vskip -0.2in
\end{figure}

In theory this increased degree of discreteness of the absolute rank distance relative to the squared one suggests that the squared rank distance might be a potentially better metric. However, in practice, this increased discreteness might not be an issue because the number of possible permutations grows extremely fast as the value of $n$ increases. Figure \ref{fig:bounded.distances.100} provides an illustration for $n = 100$. Now, due to the large possible number of permutations we only report the rank distances for 10,000 randomly sampled $\bfr_2$ permutations of the $\bfr_1^\star = (1, 2, \ldots, 100)^T$ vector (while also including the computations for $\bfr_2 = \bfr_1^\star$ and $\bfr_2 = \bfr_2^\star = (100, 99, 1)^T$ permutations). The bounded absolute rank distance (panel a) no longer shows an accentuated discreteness compared to the bounded squared rank distance (panel b).

We describe next how to obtain bounded versions of the MAE and MSE metrics for measuring information loss. Consider the rank-based version of the MAE information loss metric,
\begin{equation}
\mbox{rMAE} = \frac{1}{n \, p} \sum_{j = 1}^p \sum_{i = 1}^n |r_{ij} - \tilde{r}_{ij}|~.
\label{eq:rmae}
\end{equation}
This metric will achieve its maximum value when we replace the $\sum_{i = 1}^n |r_{ij} - \tilde{r}_{ij}|$ term by $2 \, \sum_{k = 1}^{K} (n - 2 \, k + 1)$, where $K$ equals $n/2$ if $n$ is even and $(n-1)/2$ if $n$ is odd. Therefore, the maximum value that rMAE can achieve is,
\begin{equation}
\max\{\mbox{rMAE}\} = \frac{1}{n \, p} \sum_{j = 1}^p 2 \, \sum_{k = 1}^{K} (n - 2 \, k + 1) = \frac{2}{n} \sum_{k = 1}^{K} (n - 2 \, k + 1)~.
\label{eq:max.rmae}
\end{equation}
We can then define a bounded version of the rank-based MAE metric, with range in the [0, 1] interval by dividing equation (\ref{eq:rmae}) by equation (\ref{eq:max.rmae}),
\begin{equation}
\mbox{brMAE} = \frac{1}{2 \, p \, \sum_{k = 1}^{K} (n - 2 \, k + 1)} \sum_{j = 1}^p \sum_{i = 1}^n |r_{ij} - \tilde{r}_{ij}|~.
\end{equation}

By a similar rationale the bounded version of the rank-based MSE metric is given by,
\begin{equation}
\mbox{brMSE} = \frac{1}{2 \, p \, \sum_{k = 1}^{K} (n - 2 \, k + 1)^2} \sum_{j = 1}^p \sum_{i = 1}^n (r_{ij} - \tilde{r}_{ij})^2~.
\end{equation}

\section{Empirical comparisons}

We compared the brMAE and brMSE metrics against the MAE, MSE, IL1, and IL1s metrics using microaggregation~\cite{DomingoFerrerMateoSanz2002}, noise addition~\cite{Brand2004}, and rank swapping~\cite{Moore1996} perturbation methods applied to the Tarragona and Census datasets (where the Census dataset is denoted as CASCrefmicrodata.). For the microaggregation, we adopted the maximum distance to average vector (mdav) method~\cite{DomingoFerrerMateoSanz2002}, and a grid of aggregation parameters given by,
\begin{equation}
\mbox{aggregation parameter grid} = \{1, 2, \ldots, 30 \}~.
\label{eq:aggr.par.grid}
\end{equation}
For the noise addition method we performed experiments using independent additive noise and correlated noise~\cite{Brand2004} over the grid of noise percentage parameters,
\begin{equation}
\mbox{noise \% parameter grid} = \{1, 2, \ldots, 300 \}~.
\label{eq:noise.par.grid}
\end{equation}
For the rank swapping~\cite{Moore1996} approach, we adopted the grid of percentage parameters ($P$),
\begin{equation}
\mbox{$P$ parameter grid} = \{0.001, 0.002, \ldots, 0.3 \}~.
\label{eq:rankswap.par.grid}
\end{equation}
All analysis were performed using the sdcMicro R package~\cite{sdcMicro2015}.

For each perturbation method and each dataset we: (i) produced a series of masked datasets generated by the application of increasing amounts of perturbation, as described in the perturbation parameter grids in (\ref{eq:aggr.par.grid}), (\ref{eq:noise.par.grid}), and (\ref{eq:rankswap.par.grid}); (ii) computed the brMAE, brMSE, MAE, MSE, IL1, and IL1s metrics for each perturbed dataset; (iii) computed Spearman correlations between the brMAE, brMSE, MAE, MSE, IL1, and IL1s metrics against the grid of perturbation parameters (tuning parameters) adopted for each perturbation method; and (iv) computed Spearman correlations between the brMAE (brMSE) metrics against the MAE, MSE, IL1, and IL1s metrics.

\begin{figure}[!h]
\centerline{\includegraphics[width=4.7in]{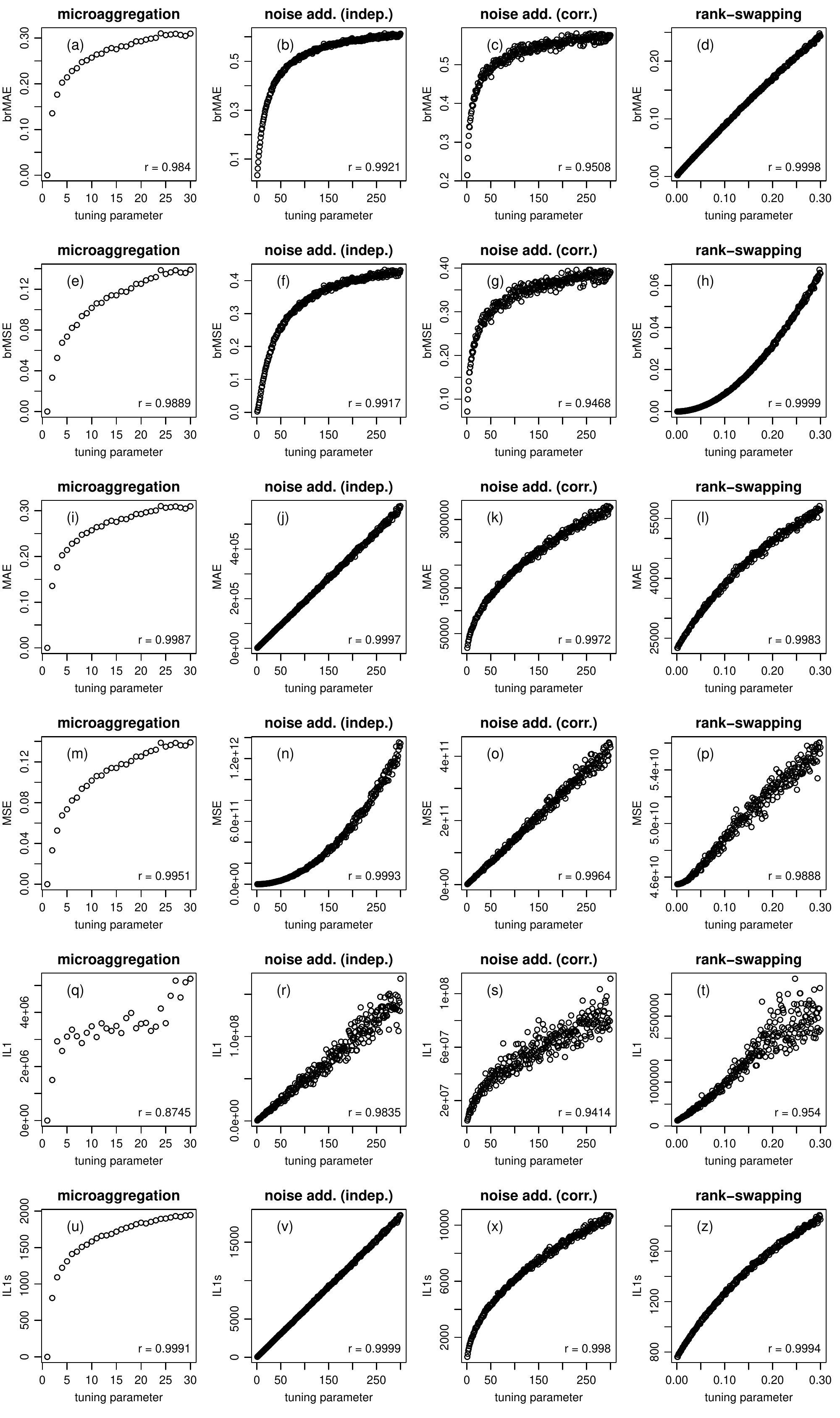}}
\vskip -0.15in
\caption{Scatterplots of the metrics versus the tuning (perturbation) parameter grids across all information loss metrics and perturbation methods in the Tarragona dataset.}
\label{fig:tuning.tarragona}
%\vskip -0.2in
\end{figure}

\begin{figure}[!h]
\centerline{\includegraphics[width=4.7in]{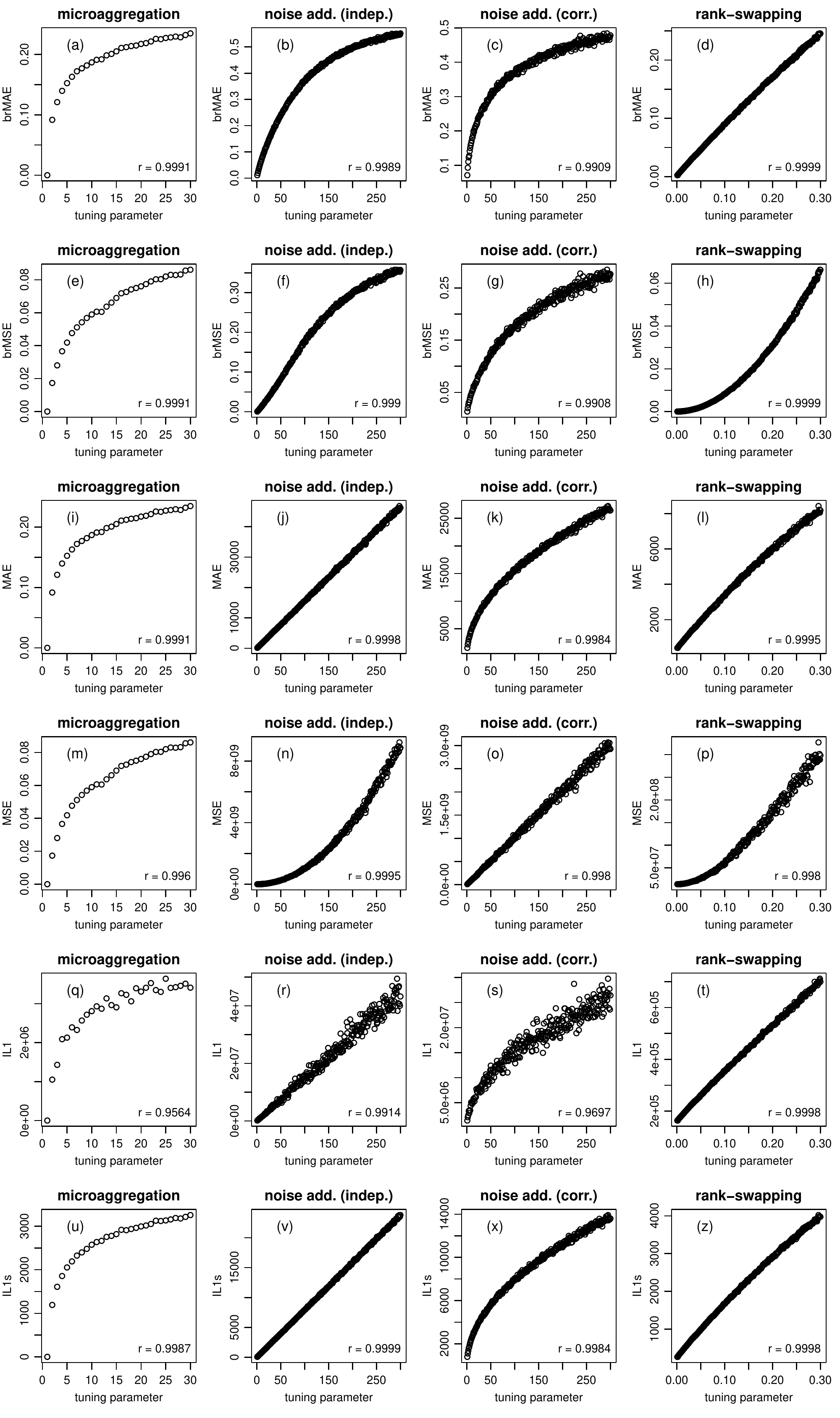}}
\vskip -0.15in
\caption{Scatterplots of the metrics versus the tuning (perturbation) parameter grids across all information loss metrics and perturbation methods in the Census dataset.}
\label{fig:tuning.census}
%\vskip -0.2in
\end{figure}

\begin{figure}[!h]
\centerline{\includegraphics[width=\linewidth]{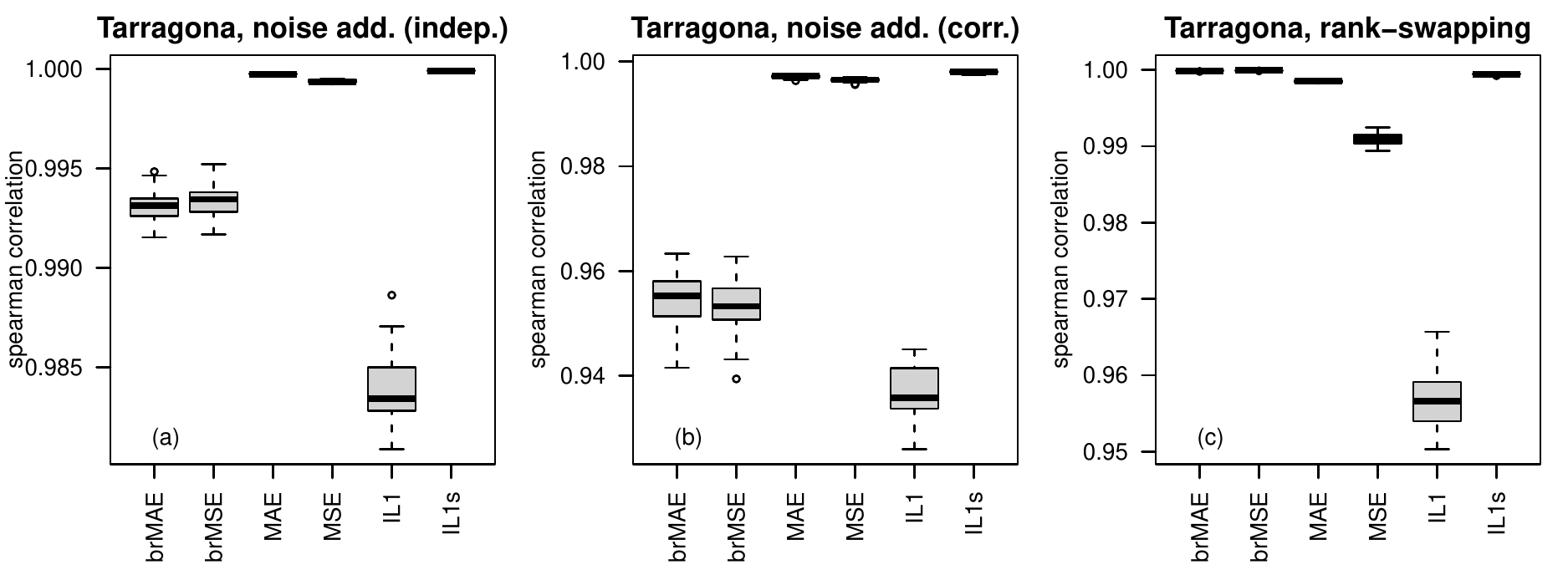}}
\vskip -0.15in
\caption{Spearman correlations between the metrics and the perturbation parameter grids across 30 replications of the perturbations in the Tarragona dataset.}
\label{fig:replications.tarragona}
%\vskip -0.2in
\end{figure}

\begin{figure}[!h]
\centerline{\includegraphics[width=\linewidth]{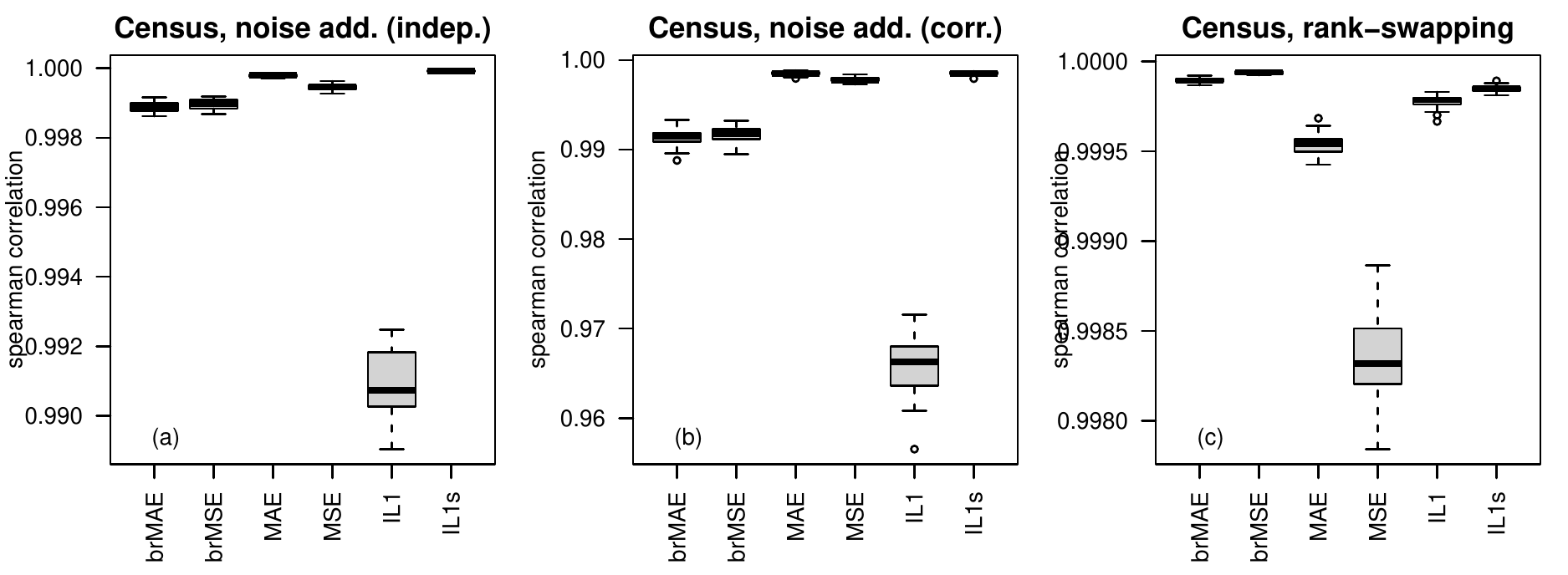}}
\vskip -0.15in
\caption{Spearman correlations between the metrics and the perturbation parameter grids across 30 replications of the perturbations in the Census dataset.}
\label{fig:replications.census}
%\vskip -0.2in
\end{figure}

Figures \ref{fig:tuning.tarragona} and \ref{fig:tuning.census} show scatterplots of the estimated information loss metrics versus the grid of perturbation parameters for all 6 metrics and 4 perturbation methods in the Tarragona and Census datasets, respectively. As expected, for all perturbation methods the amount of information loss tended to increase with the amount of perturbation applied to the datasets. The plots also report the Spearman correlations ($r$), which tended to be quite high. Due to the random nature of the noise-addition and rank-swapping methods, we report in Figures \ref{fig:replications.tarragona} and \ref{fig:replications.census} the distributions of the Spearman correlations for 30 distinct replications of the perturbation experiments in the Tarragona and Census datasets, respectively. Note that while the boxplots show systematic differences between the methods (with the IL1 metric tending to show the weakest associations with the perturbation parameter grids, the IL1s metric outperforming the other ones in the noise-addition experiments, and the brMAE and brMSE metrics outperforming the others on the rank-swapping experiments), these differences were quite small, with the Spearman correlations being quite high for all metrics (note the scales in the y-axes).

\begin{figure}[!h]
\centerline{\includegraphics[width=\linewidth]{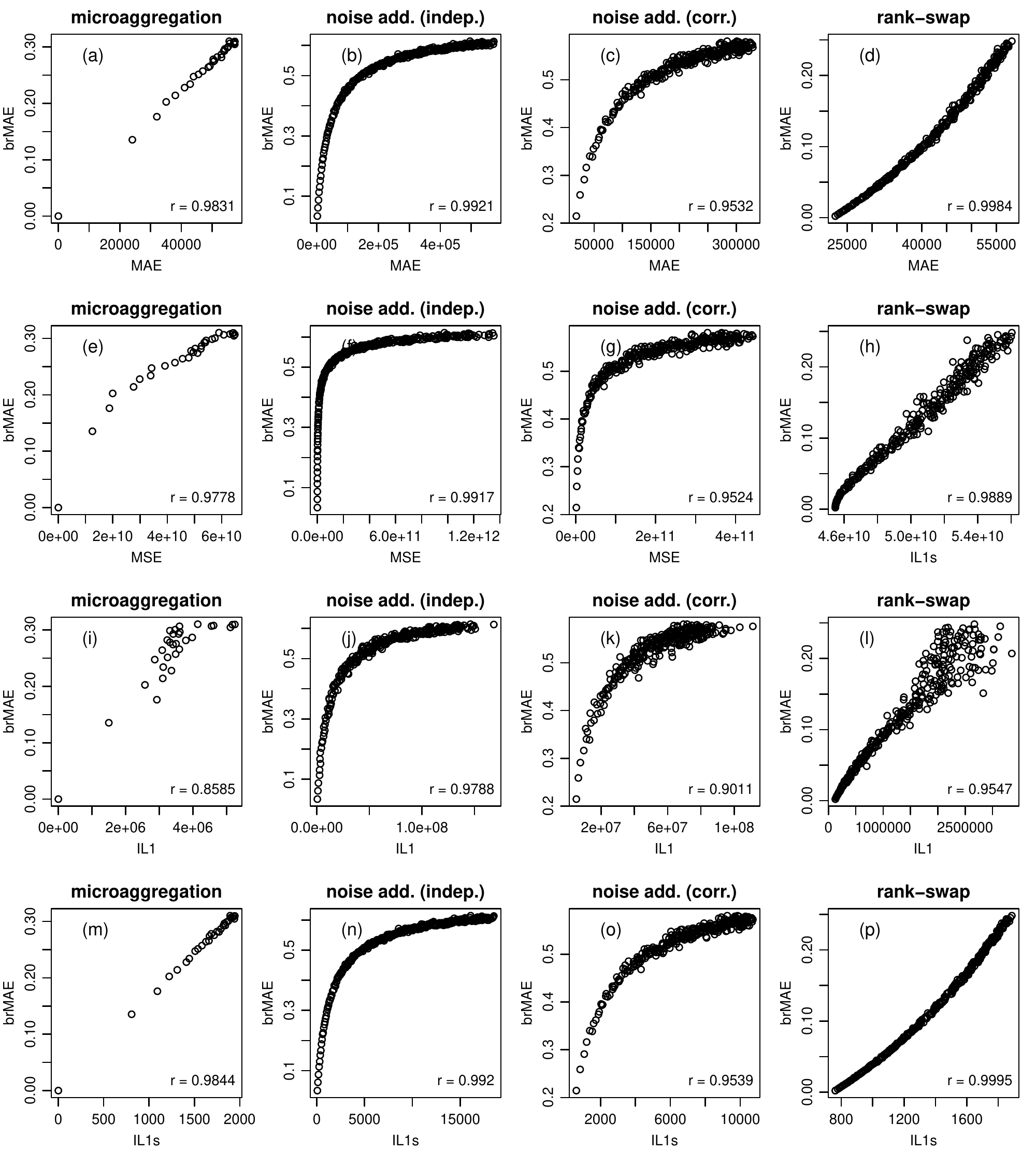}}
\vskip -0.15in
\caption{Scatterplots of brMAE against the MAE, MSE, IL1, and IL1s information loss metrics across all perturbation methods based on the Tarragona dataset.}
\label{fig:brmae.tarragona}
%\vskip -0.2in
\end{figure}

\begin{figure}[!h]
\centerline{\includegraphics[width=\linewidth]{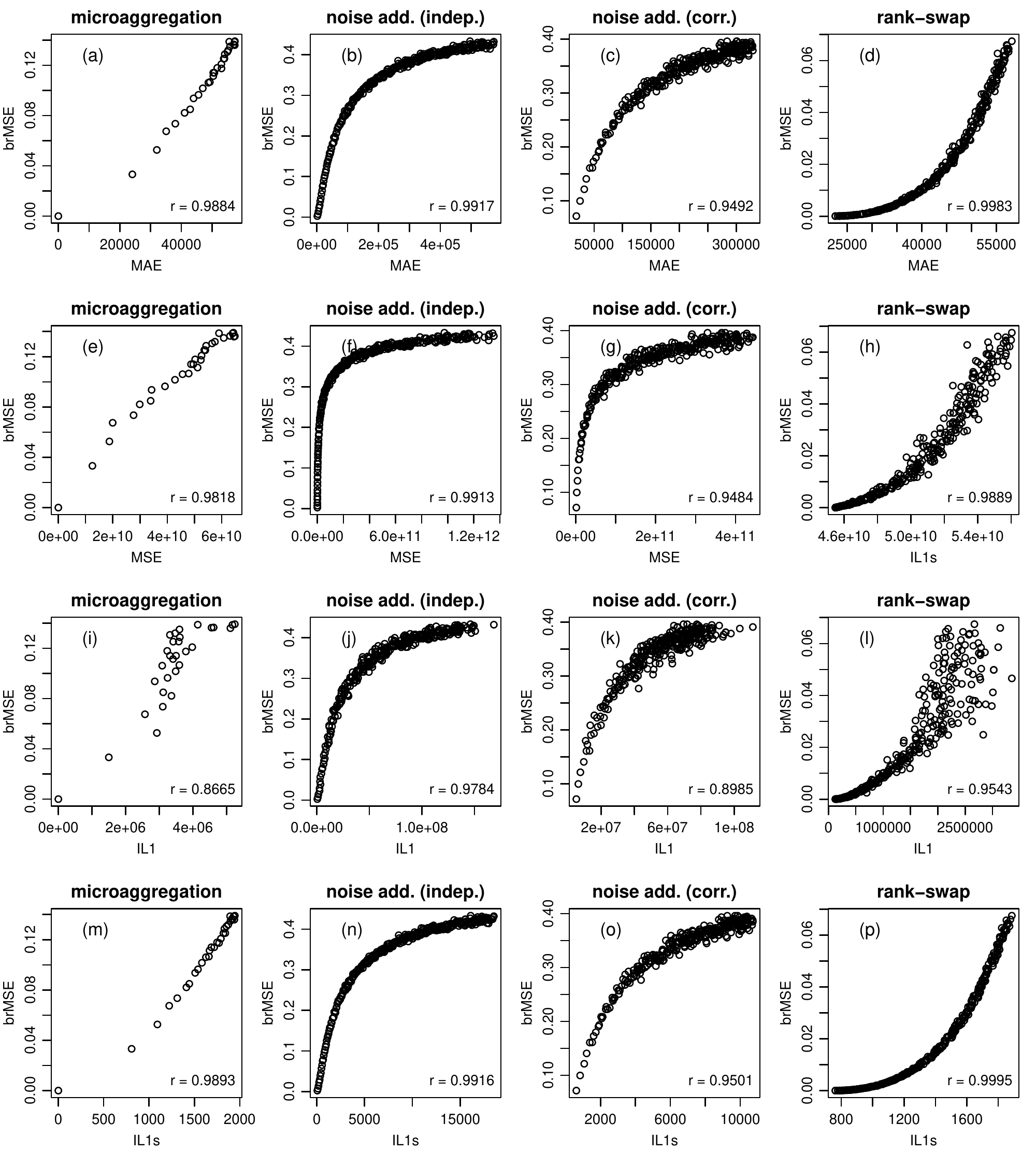}}
\vskip -0.15in
\caption{Scatterplots of brMSE against the MAE, MSE, IL1, and IL1s information loss metrics across all perturbation methods based on the Tarragona dataset.}
\label{fig:brmse.tarragona}
%\vskip -0.2in
\end{figure}

\begin{figure}[!h]
\centerline{\includegraphics[width=\linewidth]{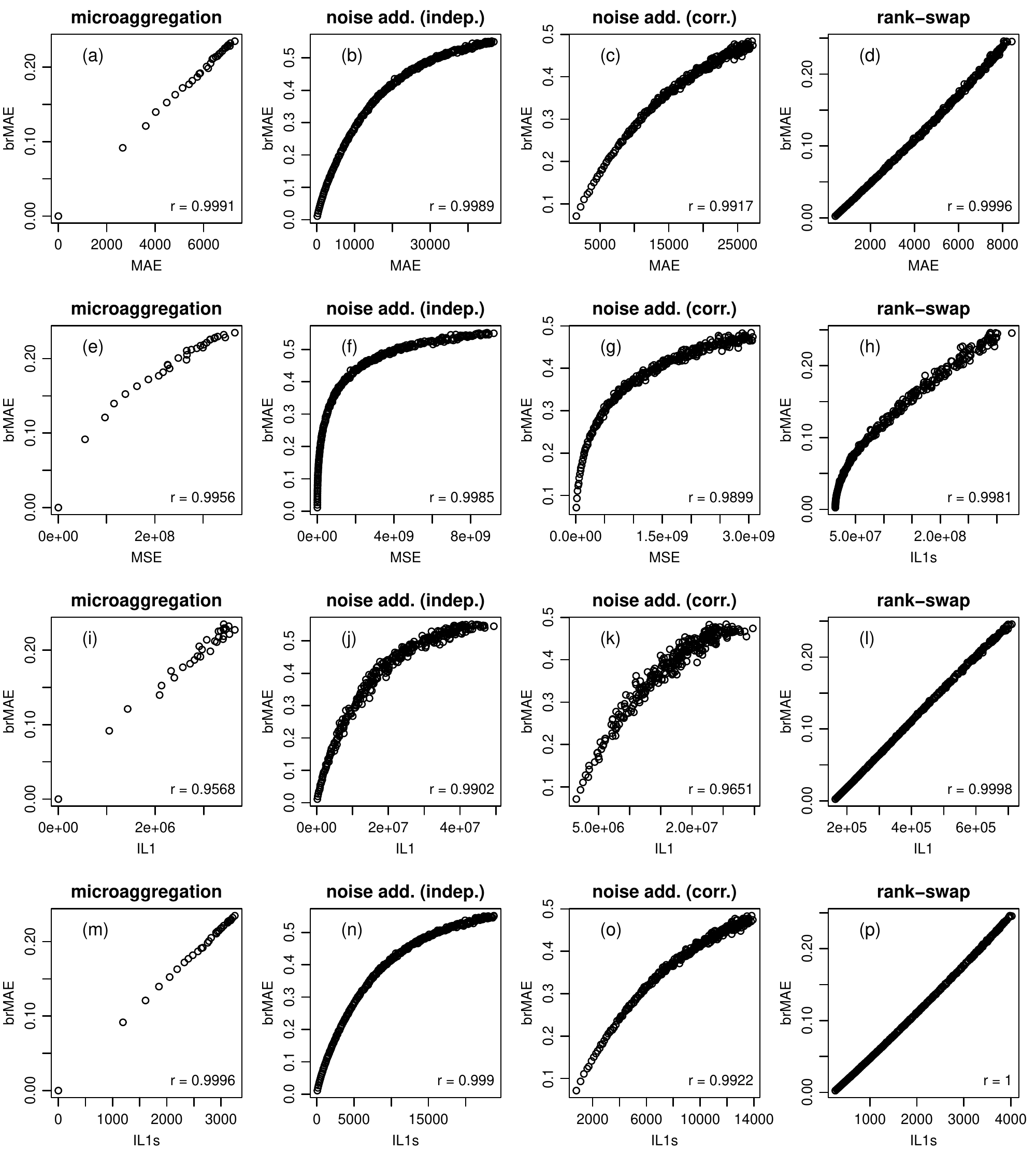}}
\vskip -0.15in
\caption{Scatterplots of brMAE against the MAE, MSE, IL1, and IL1s information loss metrics across all perturbation methods based on the Census dataset.}
\label{fig:brmae.census}
%\vskip -0.2in
\end{figure}

\begin{figure}[!h]
\centerline{\includegraphics[width=\linewidth]{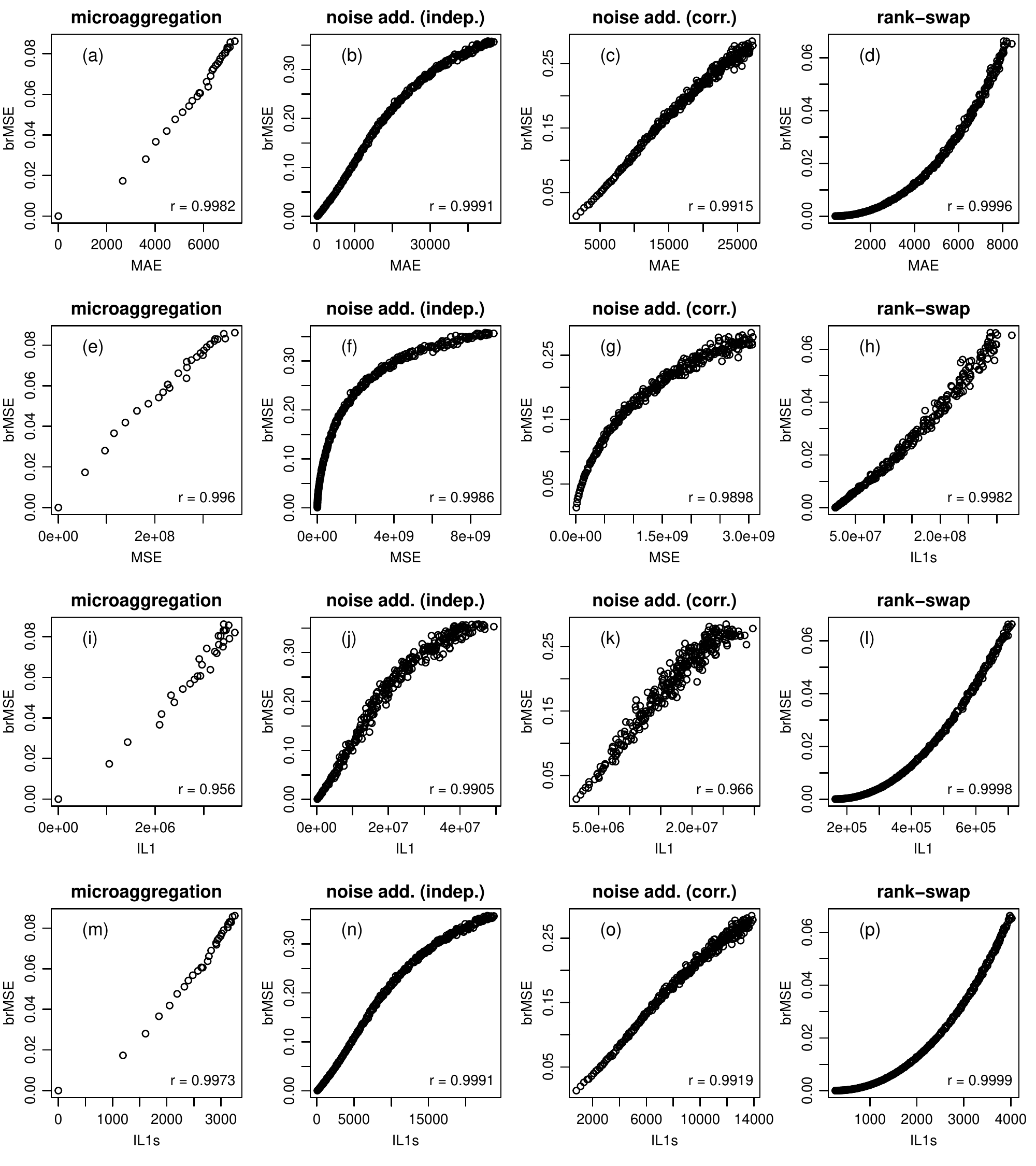}}
\vskip -0.15in
\caption{Scatterplots of brMSE against the MAE, MSE, IL1, and IL1s information loss metrics across all perturbation methods based on the Census dataset.}
\label{fig:brmse.census}
%\vskip -0.2in
\end{figure}

Figure \ref{fig:brmae.tarragona} reports the results for the comparison of brMAE against the MAE, MSE, IL1, and IL1s metrics in the Tarragona dataset. Panels a to d show the comparisons against MAE across all four perturbation methods, while panels e to h, panels i to l, and panels m to p show the comparisons against MSE, IL1, and IL1s, respectively. Figure \ref{fig:brmse.tarragona} report analogous comparisons for the brMSE metric, while Figures \ref{fig:brmae.census} and Figure \ref{fig:brmse.census} report analogous comparisons based on the Census data. In all comparisons, we observe a strong association between the rankings obtained with the bounded and unbounded metrics.

\section{Discussion}

In this note we propose bounded rank-based versions of the MAE and MSE distance based information loss metrics. The proposed metrics are bounded between 0 and 1 aiming to facilitate comparisons with disclosure risk metrics. Despite the fact that we incur some loss of information when we work with ranks rather than the raw numeric values, our empirical evaluations suggest that the brMAE and brMSE metrics can be used in place of (or in conjunction to) the current unbounded distance-based IL metrics, when performing empirical evaluations of SDC methods.


\begin{thebibliography}{50}

\bibitem{Brand2004} Brand R (2004). Microdata protection through noise addition. In Privacy in Statistical Databases. Lecture Notes in Computer Science. Springer, pages 347-359.

\bibitem{DomingoFerrerMateoSanzTorra2001} Domingo-Ferrer, J., Mateo-Sanz, J.M., and Torra, V. 2001. Comparing sdc methods for microdata on the basis of information loss and disclosure risk. In Pre-proceedings of ETK-NTTS'2001 vol. 2, Luxemburg: Eurostat, pp. 807-826

\bibitem{DomingoFerrerTorra2001} Domingo-Ferrer, J. and Torra, V. (2001) A quantitative comparison of disclosure control methods for microdata. In Confidentiality,  disclosure,  and  data  access: theory and practical applications for statistical agencies, edited by P. Doyle, J. Lane, J. Theeuwes, and L. Zayatz.  111-133.  Elsevier.

\bibitem{DomingoFerrerTorra2001b} Domingo-Ferrer, J. and Torra, V. 2001a. Disclosure protection methods and information loss for microdata. In Confidentiality, Disclosure and Data Access: Theory and Practical Applications for Statistical Agencies, P. Doyle, J.I. Lane, J.J.M. Theeuwes, and L. Zayatz (Eds.), North-Holland: Amsterdam, pp. 91-110, http://vneumann.etse.urv.es/publications/bcpi

\bibitem{DomingoFerrerMateoSanz2002} Domingo-Ferrer J and Mateo-Sanz JM (2002). Practical data-oriented microaggregation for statistical disclosure control. IEEE Trans. on Knowledge and Data Engineering, 14(1):189-201.

\bibitem{DomingoFerrerMuralidharBrasAmoros2020} Domingo-Ferrer, J., Muralidhar, K., Bras-Amoros, M. (2020) General confidentiality and utility metrics for privacy-preserving data publishing based on the permutation model. IEEE Transactions on Dependable and Secure Computing, 18(5):2506-2517. DOI: 10.1109/TDSC.2020.2968027.

\bibitem{Drechsler2011} Drechsler, J. (2011). Synthetic Data Sets for Statistical Disclosure Control. Springer, New York.

\bibitem{hundepool2012} Hundepool, A., Domingo-Ferrer, J., Franconi, L., Giessing, S., Nordholt, E. S., Spicer, K., de Wolf P. (2012) Statistical Disclosure Control, Wiley.

\bibitem{PilPaper2005} Mateo-Sanz, J.M., Domingo-Ferrer, J., Sebe F. (2005) Probabilistic information loss measures in confidentiality protection of continuous microdata. Data Mining and Knowledge Discovery 11(2):181-193.

\bibitem{MateoSanzDomingoSebeFerrer2004} Mateo-Sanz, J.M., Sebe, F., Domingo-Ferrer, J. (2004) Outlier protection in continuous microdata masking. International Workshop on Privacy in Statistical Databases. PSD 2004: Privacy in Statistical Databases pp 201-215.

\bibitem{Moore1996} Moore, Jr R (1996). Controlled data-swapping techniques for masking public use microdata, U.S. Bureau of the Census Statistical Research Division Report Series, RR 96-04.

\bibitem{PagliucaSeri1999} Pagliuca, D., and G. Seri (1999) Some results of individual ranking method on the system of enterprise accounts annual survey, Esprit SDC Project, Deliverable MI-3/D2.

\bibitem{RProject2022} R Core Team (2022). R: A language and environment for statistical computing. R Foundation for Statistical Computing, Vienna, Austria. URL https://www.R-project.org/.

\bibitem{Templ2006} Templ M (2006). Software development for SDC in R. In: Domingo-Ferrer J, Franconi L (eds.) PSD 2006. LNCS, vol. 4302, pp. 347-359. Springer, Heidelberg.

\bibitem{sdcMicro2015} Templ M, Kowarik A, and Meindl B (2015). Statistical disclosure control for micro-data using the R package sdcMicro. Journal of Statistical Software, 67(4), 1-36. https://doi.org/10.18637/jss.v067.i04

\bibitem{Woo2009} Woo, M., J. Reiter, A. Oganian, and A. Karr. (2009) Global measures of data utility for microdata masked for disclosure limitation. Journal of Privacy and Confidentiality, 1: 111-124.

\bibitem{Yancey2002} Yancey, W.E., Winkler, W.E., and Creecy, R.H. 2002. Disclosure risk assessment in perturbative microdata protection. In Inference Control in Statistical Databases, J. Domingo-Ferrer (Ed.), volume 2316 of LNCS, Berlin, Heidelberg: Springer, pp. 135-152

\end{thebibliography}
\end{document}